\documentstyle[twoside,fleqn,epsf,espcrc2]{article}

\newcommand{\nc}{\newcommand}

\nc{\tkl}{{{\small\sf T}{$\chi$}{\small\sf L}} }
\nc{\tkll}{{{\sf T}{$\chi$}{\sf L}}}
\nc{\sesaml}{{\sf SESAM}}
\nc{\sesam}{{\small\sf SESAM }}
\nc{\sesams}{{\small\sf SESAM}'s }
\nc{\sesamc}{{\small\sf SESAM} collaboration}
\nc{\largepspicture}[1]{\centerline{\setlength\epsfxsize{10cm}\epsfbox{#1}}}
\nc{\vlargepspicture}[1]{\centerline{\setlength\epsfxsize{15cm}\epsfbox{#1}}}
\nc{\smallpspicture}[1]{\centerline{\setlength\epsfxsize{8.0cm}\epsfbox{#1}}}
\nc{\tinypspicture}[1]{\centerline{\setlength\epsfxsize{7.0cm}\epsfbox{#1}}}
\nc{\ewhxy}[3]{\setlength{\epsfxsize}{#2}
            \setlength{\epsfysize}{#3}\epsfbox[0 20 660 580]{#1}}
\nc{\ewxy}[2]{\setlength{\epsfxsize}{#2}\epsfbox[10 30 640  590]{#1}}
\nc{\ewxynarrow}[2]{\setlength{\epsfxsize}{#2}\epsfbox[10 30 560 590]{#1}}
\nc{\ewxyvnarrow}[2]{\setlength{\epsfxsize}{#2}\epsfbox[10 30 520 590]{#1}}
\nc{\ewxywide}[2]{\setlength{\epsfxsize}{#2}\epsfbox[0 20 380 590]{#1}}
\nc{\err}[2]{\raisebox{0.08em}{\scriptsize{$\hspace{-0.8em}\begin{array}{@{}l@{}}
                     \plus\makebox[0.55em][r]{#1}\\[-0.15em]
                     \minus\makebox[0.55em][r]{#2}
                     \end{array}$}}}
\nc{\plus}{\makebox[15pt][c]{$+$}}
\nc{\minus}{\makebox[15pt][c]{$-$}}
\nc{\mpr}{\frac{m_{\pi}}{m_{\rho}}}
\nc{\beq}{\begin{equation}}
\nc{\eeq}{\end{equation}}
\nc{\prd}{Phys.~Rev.~D}
\nc{\prl}{Phys.~Rev.~Lett.}
\nc{\plb}{Phys.~Lett.~B}

\nc{\nn}{\nonumber}
\nc{\pr}{\prime}
\nc{\bea}{\begin{eqnarray}}
\nc{\eea}{\end{eqnarray}}
\nc{\ba}{\begin{array}}
\nc{\ea}{\end{array}}
\nc{\bt}{\begin{tabular}}
\nc{\et}{\end{tabular}}
\nc{\bc}{\begin{center}}
\nc{\ec}{\end{center}}

\nc{\Tr}{{\rm{Tr}}}
\nc{\muco}{\multicolumn}
\nc{\sss}{\scriptscriptstyle}

\nc{\amsbar}{\alpha_{\overline{\rm{MS}}}}
\nc{\co}{{\cal O}}
\nc{\ce}{{\cal E}}

\title{
\hfill\begin{minipage}{0pt}\scriptsize\vspace*{-1.5cm} \begin{tabbing}
\hspace*{\fill} HLRZ1998-55\\ 
\hspace*{\fill} WUP-TH 98-30 
\end{tabbing} 
\end{minipage}\\[-8pt]
An Estimate of $\alpha_S$ from Bottomonium in Unquenched
  QCD}

\author{
  A.~Spitz for the \sesam Collaboration$^{\rm a,b}$ \\[6pt]
  {\small  {\rm $^a$}HLRZ c/o Forschungszentrum J\"ulich, D-52425 J\"ulich,
  and DESY, D-22603 Hamburg, Germany,\\
  {\rm $^b$}Physics Department, University of Wuppertal, D-42097
  Wuppertal, Germany.}}

\begin{document}

\begin{abstract}
We estimate the strong coupling constant from the perturbative
expansion of the plaquette. The scale is set by the 2S-1S and 1P-1S
splittings in bottomonium which are computed in NRQCD on dynamical
gauge configurations with $n_f=2$ degenerate Wilson quarks at
intermediate masses. We have increased the statistics of our spectrum 
calculation in order to reliably extrapolate in the sea-quark mass. We
find a value of $\amsbar (m_Z) = 0.1118(26)$ which is somewhat lower
than previous estimates within NRQCD.  
\end{abstract}

\maketitle

%
%
\section{Introduction}
In recent years lattice calculations of the strong coupling constant 
have achieved a remarkable precision. This is mainly due to the
availability of gauge configurations that partially include the
effects of vacuum polarisation. A promising approach has been
pioneered by the NRQCD Collaboration \cite{Davies_alphas}. They
examine expectation values of small Wilson loops that are very easily
and accurately computed on the lattice and for which perturbative
expansions are known to $O(\alpha_S^2)$. The preferred choice to fix
the lattice scale, $a^{-1}$ in this context are the lowest radial and
orbital splittings in the $\Upsilon$ system since they do not require
a precise tuning of the b-quark mass, $m_b$ and are expected to exhibit
little dependence on the light quark masses. In this short note we
present a determination of $\alpha_S$ along the lines of
Ref.\cite{Davies_alphas} that is able to clarify two important
issues: first, we use dynamical Wilson quarks instead of the staggered
formulation, so that one may estimate the impact of different
discretisation schemes. Second, we calculate the dependence of
$\Upsilon$ levels on the dynamical quark mass, $m_q$ and thus reduce the
systematic error from simulating overly heavy sea-quarks.  
%
%
\section{Simulation}
\begin{table}
\begin{center}
\begin{tabular}{llll}
\hline\hline
\multicolumn{4}{l}{Dynamical Wilson $\beta=5.6$ , $16^3\times 32$} \\
\hline
$\kappa$           & 0.1560 & 0.1570 & 0.1575 \\
$m_{\pi}/m_{\rho}$ & 0.83   & 0.76   & 0.69   \\
configurations     & 206    & 192    & 203    \\
measurements       & 824    & 2496   & 2436   \\
\hline\hline
\end{tabular}
\vspace{0.2cm}
\caption{\sl Simulation set-up. \label{tab:simsetup}}
\vspace{-1.3cm}
\end{center}
\end{table}
We employ NRQCD at next-to-leading order, i.e. spin-independent
operators of $O(m_b v^2)$, $O(m_b v^4)$ and spin-dependent terms of
$O(m_b v^4)$ and $O(m_b v^6)$ have been included into the action. We
rely on the recipe of tadpole improvement using $u_0$ from the
mean-link in Landau gauge to account for the bulk of loop
contributions and otherwise stick to tree-level couplings. Propagators
are evaluated on SESAM's gauge configurations with two degenerate
flavours of Wilson quarks \cite{SESAM_review}. We exploit each configuration
more than once starting at different sources, see Table \ref{tab:simsetup}.  
We did not tune the bare b-quark mass but kept $am_b = 1.7$ throughout
the simulation. This value reproduces the correct $\Upsilon$ mass in
the quenched approximation and it turns out to be adequate in
the full theory, too, leading to kinetic masses $m_{\rm kin}(\Upsilon) =
9.97(28),9.63(24),9.68(27)\; {\rm GeV}$ for $\kappa =
0.1560,0.1570,0.1575$, respectively. Compared to the results presented
in \cite{SESAM_nrqcd}, we have significantly increased our statistics
except for the heaviest sea-quark mass. 
%
%
\section{Bottomonium Spectroscopy}
The results quoted here are taken from simultaneous multi-exponential 
fits to correlators with a local sink and a source that is smeared
with a carefully chosen potential model wave function. These
correlators have the cleanest signals and lead to very stable
fits. Following \cite{Shigemitsu_lat96} we 
extrapolate the energy splittings linearly in the sea-quark mass,
\begin{equation}
    a\Delta E = a\Delta E_0 + c\, am_q 
\end{equation}   
to $am_{\rm eff} \equiv am_s/3 \sim 0.0159$ which is somewhat below
our lightest quark mass. Fit parameters are listed in Table
\ref{tab:extrap_results} together with the energy splittings at
$m_{\rm eff}$. We find the dependence on the dynamical quark
mass to be much smaller than in the light hadron sector. 
\begin{table}
\begin{center}
\begin{tabular}{clll}
\hline\hline
splitting & $a\Delta E_0$ & $c$ & \hspace{-0.1cm}$a\Delta E(m_s/3)$ \\
\hline
$2^1S_0 - 1^1S_0$ & 0.209(21) & 1.2(7)   & 0.229(10) \\
$2^3S_1 - 1^3S_1$ & 0.209(18) & 1.1(7)   & 0.226(9)  \\
$1^3\bar P - 1^3S_1$ & 0.163(9)  & 0.4(3)   & 0.170(5)  \\
$2^1P_1 - 1^1P_1$ & 0.152(24) & 1.8(7)   & 0.181(15) \\
\hline\hline
\end{tabular}
\vspace{0.1cm}
\caption{\sl Results of the extrapolation in the sea-quark
  mass. $^3\bar P$ is the spin-averaged triplet-P
  state. \label{tab:extrap_results}}  
\vspace{-1.5cm}
\end{center}
\end{table} 
\begin{table}
\begin{center}
\begin{tabular}{llll}
\hline\hline
& \multicolumn{2}{c}{$a^{-1}[{\sf GeV}]$} & $R_{\sf SP}$ \\
& $\Upsilon ' - \Upsilon$ & $\bar\chi - \Upsilon$ & \\
\hline
$n_f = 0,\beta = 6.0$ & 2.29(11) & 2.68(9) & 1.49(10) \\
\hline
$\kappa = 0.1560 $ & 2.12(16) & 2.38(8) & 1.44(10) \\
\hline 
$\kappa = 0.1570 $ & 2.34(9) & 2.50(8) & 1.37(7) \\
\hline 
$\kappa = 0.1575 $ & 2.43(8) & 2.57(7) & 1.35(6) \\
\hline
$ m_s/3 $ & 2.49(10) & 2.59(7) & 1.33(6) \\
\hline\hline
\end{tabular}
\vspace{0.1cm}
\caption[Lattice spacings]{\sl Determination of the lattice spacing
  from the $2^3S_1-1^3S_1$ and $1^3\bar P-1^3S_1$ splittings. Their
  ratio, $R_{\sf SP}$ is to be compared to the experimental value of
  1.28.\label{tab:spacings}} 
\vspace{-1.0cm}
\end{center}
\end{table}
\begin{figure}
  \smallpspicture{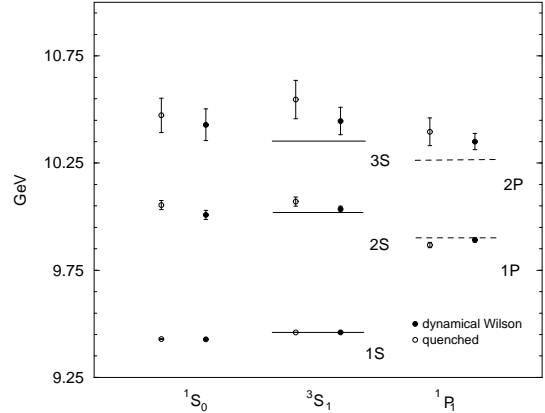}
\vspace{-1.5cm}
\caption{\sl Bottomonium spectrum -- radial level splittings. The
  $^3S_1$ ground state is fixed to the experimental $\Upsilon$
  mass. Open symbols denote quenched results, filled symbols $n_f = 2$
  results at $m_q = m_s/3$. Solid lines mark experimental values,
  dashed lines the positions of the spin-averaged $^3P_J$ states.
  \label{fig:bb_spectrum}} 
\vspace{-0.8cm}
\end{figure}
Results for lattice spacings are summarised in Table
\ref{tab:spacings}. We use the average spacing from $2^3S_1-1^3S_1$
and $1^3\bar P-1^3S_1$ (at $m_q = m_{\rm eff}$ in the unquenched case) to
convert our results to physical units. Note that the average $a^{-1}$ for
$\beta=5.6$, $n_f=2$ at $am_{\rm eff}$ agrees well with the quenched
one at $\beta = 6.0$, so that we can directly compare both theories.
Figure \ref{fig:bb_spectrum} collects our spectrum results for two and zero
flavours. It is obvious that the gross level structure computed on quenched
configurations disagrees with experiment whereas the $n_f=2$ results
are in much closer agreement. This is also
evident from Table \ref{tab:spacings} which shows that the 
lattice spacings from the 2S-1S and the 1P-1S splittings do not
agree in the quenched theory whereas they coincide when two
dynamical quarks are switched on. 

\section{Plaquette Coupling}
\begin{figure}
\begin{center}
  \vspace{-0.5cm}
  \smallpspicture{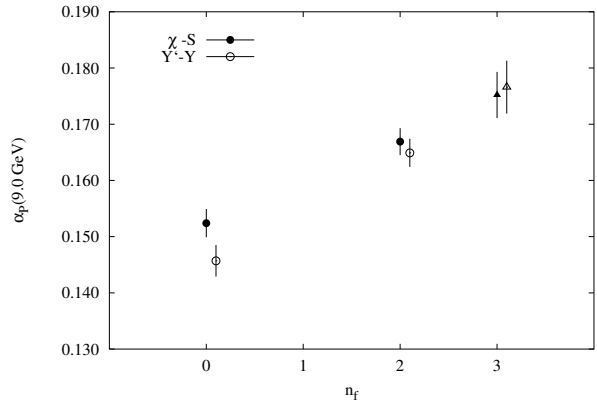}
\vspace{-1.0cm}
\caption{\sl The plaquette coupling $\alpha_P$ as a function of the number
  of dynamical flavours. The triangles result from an extrapolation in
  the inverse flavour number. \label{fig:alphaP_nf}}
\vspace{-1.0cm}
\end{center}
\end{figure}

We adopt the definition of the plaquette coupling given in
\cite{Davies_alphas}, 
\bea
\label{eq:def_alpha_P}
  \lefteqn{-\ln W_{11} \equiv -\ln\langle \frac{1}{3}\Re\Tr\Box\rangle
  =} \nonumber \\
  && \!\!\!\!\!\!\!\frac{4\pi}{3}\alpha_P\left(\frac{3.41}{a}\right)\left[ 1 -
  \left(1.1855 + 0.0249\; n_f\right)\alpha_P\right] \;
\eea
which equals the expansion in $\alpha_V$ but shifts
truncation errors to the conversion into standard
couplings later on. In Table \ref{tab:alphaP} we summarise the
couplings $\alpha_P$ obtained from Eq.(\ref{eq:def_alpha_P}) as well
as the scales determined from the 1P-1S ($\bar\chi - S$) and 2S-1S
($\Upsilon' - \Upsilon$) splittings. In the unquenched case we quote
values for both $am_q = am_s /3$ and the chiral limit to estimate the
systematic error connected to the finite sea-quark mass. Subsequently
these couplings are evolved to a common scale using the universal
two-loop $\beta$ function. The plaquette couplings in the quenched and
  unquenched theory can now 
be extrapolated to the active number of light quark flavours which is
expected to be $n_f = 3$ in the case of the low-lying $b\bar b$ bound 
states. Guided by the perturbative evolution, we extrapolate
$\alpha_P^{-1}$ linearly in $n_f$, Figure
\ref{fig:alphaP_nf}. Obviously the mismatch between 
$\alpha_P$-values obtained from different splittings in the
quenched approximation disappears, once the dynamical quarks are
switched on. 
{\tabcolsep1.0mm
\begin{table}
\begin{center}
\begin{tabular}{cllll}
\hline\hline
$M_q$ & $-\ln W_{11}$ &
$\alpha_P^{(n_f)}(\frac{3.41}{a})$ &
\multicolumn{2}{c}{$\frac{3.41}{a}$ [GeV]} \\
&&& $\bar\chi - \Upsilon$ & $\Upsilon^{\pr} - \Upsilon$ \\
\hline
0.0159   & 0.5570 & 0.1677 & 8.84(44) & 8.48(46)  \\
0        & 0.5546 & 0.1668 & 9.16(83) & 9.09(101) \\
$\infty$ & 0.5214 & 0.1518 & 9.13(50) & 7.82(54)  \\
\hline\hline 
\end{tabular}
\vspace{0.2cm}
\caption[Results for $\alpha_P\left(3.41/a\right)$]{\sl Results for
  $\alpha_P\left(3.41/a\right)$ extracted from the 
  measured plaquette values. \label{tab:alphaP}} 
\vspace{-1.0cm}
\end{center}
\end{table}
}
%
%
\section{Discussion}
To make the connection with the $\overline{MS}$-scheme one invokes
\begin{eqnarray*}
  \label{eq:conversion}
  \lefteqn{\alpha_{\overline{MS}}^{(n_f)}\left(Q\right) =} \\
  && \hspace{-0.5cm}\alpha_P^{(n_f)}\left( {\rm e}^{5/6}Q\right) \left[ 1 +
  \frac{2}{\pi}\alpha_P^{(n_f)} + C_2(n_f)\alpha_P^2 + \dots \right]
  \; ,
\end{eqnarray*}
where $C_2 = 0.96$ for $n_f = 0$. Following 
Ref. \cite{Davies_alphas}, we use this value also for $n_f=2$ and take
the whole size of the quenched two-loop contribution as an error
estimate for the conversion. 
The couplings are evolved to the Z mass applying third-order
perturbative evolution. We perform the matching at the heavy-quark
thresholds $m_c = 1.3$ GeV and $m_b = 4.1$ GeV.
We thus arrive at our final results
\begin{equation}
\alpha_{\overline{MS}}\left(m_Z\right) = \left\{\begin{array}{lll} 
    \!\!0.1118 & \!\!\! (16)(5)(20) & \!\! \bar\chi - \Upsilon \\
    \!\! 0.1124 & \!\!\!(18)(15)(20)& \!\! \Upsilon ' - \Upsilon 
    \end{array} \right. 
\end{equation}
The first error includes the statistical uncertainty and the
systematic errors due to the influence of relativistic corrections and
changes in $u_0$ on the lattice scale. The second error quantifies the
sea-quark mass dependence and the last one the truncation error in the
conversion.

We cannot confirm the staggered result \cite{Davies_alphas} but
obtain a value of $\alpha_{\overline{MS}}(m_Z)$ which is smaller by
two standard deviations. A closer inspection reveals that the effect
can be traced back to the difference between
staggered and Wilson plaquette values which is larger than anticipated
from perturbation theory, while 
the scales in both simulations are comparable. There is some
controversy about the proper way to 
determine the lattice scale in an unquenched simulation. Whereas
SESAM's strategy is to set $a^{-1}$ in the chiral limit, one may also
think of separate lattice spacings for each quark mass. A priori it is
conceivable that the value of $\alpha_P$ would be larger if it were
determined on each subensemble and then extrapolated in $m_q$. However,
we observe a very mild dependence of the $1P-1S$ splitting on the
quark mass, so that different extrapolation methods do not produce a
big effect.  
%
%

\end{document}